\documentclass[conference,10pt,twoside,twocolumn]{IEEEtran}

\usepackage{cite}
\usepackage[T1]{fontenc}
\usepackage{graphicx}
\usepackage{amssymb}
\usepackage{amsmath}
\usepackage{amsthm}
\usepackage{subfigure}
\usepackage{booktabs} 
\usepackage{multirow}
\usepackage{microtype}
\usepackage{balance}
\usepackage{halloweenmath}
\usepackage{xcolor}
\usepackage{wasysym}
\usepackage{twemojis}
\usepackage{hyperref}

\usepackage{amssymb}
\usepackage{amsfonts}
\usepackage{mathrsfs}
\usepackage{xspace}
\usepackage{bm}
\usepackage{upgreek}

\newcommand{\safemath}[2]{\newcommand{#1}{\ensuremath{#2}\xspace}}

\safemath{\bma}{\mathbf{a}}
\safemath{\bmb}{\mathbf{b}}
\safemath{\bmc}{\mathbf{c}}
\safemath{\bmd}{\mathbf{d}}
\safemath{\bme}{\mathbf{e}}
\safemath{\bmf}{\mathbf{f}}
\safemath{\bmg}{\mathbf{g}}
\safemath{\bmh}{\mathbf{h}}
\safemath{\bmi}{\mathbf{i}}
\safemath{\bmj}{\mathbf{j}}
\safemath{\bmk}{\mathbf{k}}
\safemath{\bml}{\mathbf{l}}
\safemath{\bmm}{\mathbf{m}}
\safemath{\bmn}{\mathbf{n}}
\safemath{\bmo}{\mathbf{o}}
\safemath{\bmp}{\mathbf{p}}
\safemath{\bmq}{\mathbf{q}}
\safemath{\bmr}{\mathbf{r}}
\safemath{\bms}{\mathbf{s}}
\safemath{\bmt}{\mathbf{t}}
\safemath{\bmu}{\mathbf{u}}
\safemath{\bmv}{\mathbf{v}}
\safemath{\bmw}{\mathbf{w}}
\safemath{\bmx}{\mathbf{x}}
\safemath{\bmy}{\mathbf{y}}
\safemath{\bmz}{\mathbf{z}}
\safemath{\bmzero}{\mathbf{0}}
\safemath{\bmone}{\mathbf{1}}

\bmdefine{\biad}{a}
\bmdefine{\bibd}{b}
\bmdefine{\bicd}{c}
\bmdefine{\bidd}{d}
\bmdefine{\bied}{e}
\bmdefine{\bifd}{f}
\bmdefine{\bigd}{g}
\bmdefine{\bihd}{h}
\bmdefine{\biid}{i}
\bmdefine{\bijd}{j}
\bmdefine{\bikd}{k}
\bmdefine{\bild}{l}
\bmdefine{\bimd}{m}
\bmdefine{\bind}{n}
\bmdefine{\biod}{o}
\bmdefine{\bipd}{p}
\bmdefine{\biqd}{q}
\bmdefine{\bird}{r}
\bmdefine{\bisd}{s}
\bmdefine{\bitd}{t}
\bmdefine{\biud}{u}
\bmdefine{\bivd}{v}
\bmdefine{\biwd}{w}
\bmdefine{\bixd}{x}
\bmdefine{\biyd}{y}
\bmdefine{\bizd}{z}

\bmdefine{\bixid}{\xi}
\bmdefine{\bilambdad}{\lambda}
\bmdefine{\bimud}{\mu}
\bmdefine{\bithetad}{\theta}
\bmdefine{\biphid}{\phi}
\bmdefine{\bideltad}{\delta}

\safemath{\bmia}{\biad}
\safemath{\bmib}{\bibd}
\safemath{\bmic}{\bicd}
\safemath{\bmid}{\bidd}
\safemath{\bmie}{\bied}
\safemath{\bmif}{\bifd}
\safemath{\bmig}{\bigd}
\safemath{\bmih}{\bihd}
\safemath{\bmii}{\biid}
\safemath{\bmij}{\bijd}
\safemath{\bmik}{\bikd}
\safemath{\bmil}{\bild}
\safemath{\bmim}{\bimd}
\safemath{\bmin}{\bind}
\safemath{\bmio}{\biod}
\safemath{\bmip}{\bipd}
\safemath{\bmiq}{\biqd}
\safemath{\bmir}{\bird}
\safemath{\bmis}{\bisd}
\safemath{\bmit}{\bitd}
\safemath{\bmiu}{\biud}
\safemath{\bmiv}{\bivd}
\safemath{\bmiw}{\biwd}
\safemath{\bmix}{\bixd}
\safemath{\bmiy}{\biyd}
\safemath{\bmiz}{\bizd}

\safemath{\bmxi}{\bixid}
\safemath{\bmlambda}{\bilambdad}
\safemath{\bmmu}{\bimud}
\safemath{\bmtheta}{\bithetad}
\safemath{\bmphi}{\biphid}
\safemath{\bmdelta}{\bideltad}

\safemath{\bA}{\mathbf{A}}
\safemath{\bB}{\mathbf{B}}
\safemath{\bC}{\mathbf{C}}
\safemath{\bD}{\mathbf{D}}
\safemath{\bE}{\mathbf{E}}
\safemath{\bF}{\mathbf{F}}
\safemath{\bG}{\mathbf{G}}
\safemath{\bH}{\mathbf{H}}
\safemath{\bI}{\mathbf{I}}
\safemath{\bJ}{\mathbf{J}}
\safemath{\bK}{\mathbf{K}}
\safemath{\bL}{\mathbf{L}}
\safemath{\bM}{\mathbf{M}}
\safemath{\bN}{\mathbf{N}}
\safemath{\bO}{\mathbf{O}}
\safemath{\bP}{\mathbf{P}}
\safemath{\bQ}{\mathbf{Q}}
\safemath{\bR}{\mathbf{R}}
\safemath{\bS}{\mathbf{S}}
\safemath{\bT}{\mathbf{T}}
\safemath{\bU}{\mathbf{U}}
\safemath{\bV}{\mathbf{V}}
\safemath{\bW}{\mathbf{W}}
\safemath{\bX}{\mathbf{X}}
\safemath{\bY}{\mathbf{Y}}
\safemath{\bZ}{\mathbf{Z}}

\safemath{\bZero}{\mathbf{0}}
\safemath{\bOne}{\mathbf{1}}
\safemath{\bDelta}{\mathbf{\Delta}}
\safemath{\bLambda}{\mathbf{\UpLambda}}
\safemath{\bPhi}{\mathbf{\Upphi}}
\safemath{\bSigma}{\mathbf{\Upsigma}}
\safemath{\bOmega}{\mathbf{\Upomega}}
\safemath{\bTheta}{\mathbf{\Uptheta}}

\bmdefine{\biAd}{A}
\bmdefine{\biBd}{B}
\bmdefine{\biCd}{C}
\bmdefine{\biDd}{D}
\bmdefine{\biEd}{E}
\bmdefine{\biFd}{F}
\bmdefine{\biGd}{G}
\bmdefine{\biHd}{H}
\bmdefine{\biId}{I}
\bmdefine{\biJd}{J}
\bmdefine{\biKd}{K}
\bmdefine{\biLd}{L}
\bmdefine{\biMd}{M}
\bmdefine{\biOd}{N}
\bmdefine{\biPd}{O}
\bmdefine{\biQd}{P}
\bmdefine{\biRd}{R}
\bmdefine{\biSd}{S}
\bmdefine{\biTd}{T}
\bmdefine{\biUd}{U}
\bmdefine{\biVd}{V}
\bmdefine{\biWd}{W}
\bmdefine{\biXd}{X}
\bmdefine{\biYd}{Y}
\bmdefine{\biZd}{Z}

\bmdefine{\biDelta}{\Delta}
\bmdefine{\biLambda}{\Lambda}
\bmdefine{\biPhi}{\Phi}
\bmdefine{\biSigma}{\Sigma}
\bmdefine{\biOmega}{\Omega}
\bmdefine{\biTheta}{\Theta}

\safemath{\bimA}{\biAd}
\safemath{\bimB}{\biBd}
\safemath{\bimC}{\biCd}
\safemath{\bimD}{\biDd}
\safemath{\bimE}{\biEd}
\safemath{\bimF}{\biFd}
\safemath{\bimG}{\biGd}
\safemath{\bimH}{\biHd}
\safemath{\bimI}{\biId}
\safemath{\bimJ}{\biJd}
\safemath{\bimK}{\biKd}
\safemath{\bimL}{\biLd}
\safemath{\bimM}{\biMd}
\safemath{\bimN}{\biNd}
\safemath{\bimO}{\biOd}
\safemath{\bimP}{\biPd}
\safemath{\bimQ}{\biQd}
\safemath{\bimR}{\biRd}
\safemath{\bimS}{\biSd}
\safemath{\bimT}{\biTd}
\safemath{\bimU}{\biUd}
\safemath{\bimV}{\biVd}
\safemath{\bimW}{\biWd}
\safemath{\bimX}{\biXd}
\safemath{\bimY}{\biYd}
\safemath{\bimZ}{\biZd}

\safemath{\bimDelta}{\biDelta}
\safemath{\bimLambda}{\biLambda}
\safemath{\bimPhi}{\biPhi}
\safemath{\bimSigma}{\biSigma}
\safemath{\bimOmega}{\biOmega}
\safemath{\bimTheta}{\biTheta}

\safemath{\setA}{\mathcal{A}}
\safemath{\setB}{\mathcal{B}}
\safemath{\setC}{\mathcal{C}}
\safemath{\setD}{\mathcal{D}}
\safemath{\setE}{\mathcal{E}}
\safemath{\setF}{\mathcal{F}}
\safemath{\setG}{\mathcal{G}}
\safemath{\setH}{\mathcal{H}}
\safemath{\setI}{\mathcal{I}}
\safemath{\setJ}{\mathcal{J}}
\safemath{\setK}{\mathcal{K}}
\safemath{\setL}{\mathcal{L}}
\safemath{\setM}{\mathcal{M}}
\safemath{\setN}{\mathcal{N}}
\safemath{\setO}{\mathcal{O}}
\safemath{\setP}{\mathcal{P}}
\safemath{\setQ}{\mathcal{Q}}
\safemath{\setR}{\mathcal{R}}
\safemath{\setS}{\mathcal{S}}
\safemath{\setT}{\mathcal{T}}
\safemath{\setU}{\mathcal{U}}
\safemath{\setV}{\mathcal{V}}
\safemath{\setW}{\mathcal{W}}
\safemath{\setX}{\mathcal{X}}
\safemath{\setY}{\mathcal{Y}}
\safemath{\setZ}{\mathcal{Z}}
\safemath{\emptySet}{\varnothing}

\safemath{\colA}{\mathscr{A}}
\safemath{\colB}{\mathscr{B}}
\safemath{\colC}{\mathscr{C}}
\safemath{\colD}{\mathscr{D}}
\safemath{\colE}{\mathscr{E}}
\safemath{\colF}{\mathscr{F}}
\safemath{\colG}{\mathscr{G}}
\safemath{\colH}{\mathscr{H}}
\safemath{\colI}{\mathscr{I}}
\safemath{\colJ}{\mathscr{J}}
\safemath{\colK}{\mathscr{K}}
\safemath{\colL}{\mathscr{L}}
\safemath{\colM}{\mathscr{M}}
\safemath{\colN}{\mathscr{N}}
\safemath{\colO}{\mathscr{O}}
\safemath{\colP}{\mathscr{P}}
\safemath{\colQ}{\mathscr{Q}}
\safemath{\colR}{\mathscr{R}}
\safemath{\colS}{\mathscr{S}}
\safemath{\colT}{\mathscr{T}}
\safemath{\colU}{\mathscr{U}}
\safemath{\colV}{\mathscr{V}}
\safemath{\colW}{\mathscr{W}}
\safemath{\colX}{\mathscr{X}}
\safemath{\colY}{\mathscr{Y}}
\safemath{\colZ}{\mathscr{Z}}

\safemath{\opA}{\mathbb{A}}
\safemath{\opB}{\mathbb{B}}
\safemath{\opC}{\mathbb{C}}
\safemath{\opD}{\mathbb{D}}
\safemath{\opE}{\mathbb{E}}
\safemath{\opF}{\mathbb{F}}
\safemath{\opG}{\mathbb{G}}
\safemath{\opH}{\mathbb{H}}
\safemath{\opI}{\mathbb{I}}
\safemath{\opJ}{\mathbb{J}}
\safemath{\opK}{\mathbb{K}}
\safemath{\opL}{\mathbb{L}}
\safemath{\opM}{\mathbb{M}}
\safemath{\opN}{\mathbb{N}}
\safemath{\opO}{\mathbb{O}}
\safemath{\opP}{\mathbb{P}}
\safemath{\opQ}{\mathbb{Q}}
\safemath{\opR}{\mathbb{R}}
\safemath{\opS}{\mathbb{S}}
\safemath{\opT}{\mathbb{T}}
\safemath{\opU}{\mathbb{U}}
\safemath{\opV}{\mathbb{V}}
\safemath{\opW}{\mathbb{W}}
\safemath{\opX}{\mathbb{X}}
\safemath{\opY}{\mathbb{Y}}
\safemath{\opZ}{\mathbb{Z}}
\safemath{\opZero}{\mathbb{O}}
\safemath{\identityop}{\opI}

\safemath{\veca}{\bma}
\safemath{\vecb}{\bmb}
\safemath{\vecc}{\bmc}
\safemath{\vecd}{\bmd}
\safemath{\vece}{\bme}
\safemath{\vecf}{\bmf}
\safemath{\vecg}{\bmg}
\safemath{\vech}{\bmh}
\safemath{\veci}{\bmi}
\safemath{\vecj}{\bmj}
\safemath{\veck}{\bmk}
\safemath{\vecl}{\bml}
\safemath{\vecm}{\bmm}
\safemath{\vecn}{\bmn}
\safemath{\veco}{\bmo}
\safemath{\vecp}{\bmp}
\safemath{\vecq}{\bmq}
\safemath{\vecr}{\bmr}
\safemath{\vecs}{\bms}
\safemath{\vect}{\bmt}
\safemath{\vecu}{\bmu}
\safemath{\vecv}{\bmv}
\safemath{\vecw}{\bmw}
\safemath{\vecx}{\bmx}
\safemath{\vecy}{\bmy}
\safemath{\vecz}{\bmz}

\safemath{\veczero}{\bmzero}
\safemath{\vecone}{\bmone}
\safemath{\vecxi}{\bmxi}
\safemath{\veclambda}{\bmlambda}
\safemath{\vecmu}{\bmmu}
\safemath{\vectheta}{\bmtheta}
\safemath{\vecphi}{\bmphi}
\safemath{\vecdelta}{\bmdelta}

\safemath{\matA}{\bA}
\safemath{\matB}{\bB}
\safemath{\matC}{\bC}
\safemath{\matD}{\bD}
\safemath{\matE}{\bE}
\safemath{\matF}{\bF}
\safemath{\matG}{\bG}
\safemath{\matH}{\bH}
\safemath{\matI}{\bI}
\safemath{\matJ}{\bJ}
\safemath{\matK}{\bK}
\safemath{\matL}{\bL}
\safemath{\matM}{\bM}
\safemath{\matN}{\bN}
\safemath{\matO}{\bO}
\safemath{\matP}{\bP}
\safemath{\matQ}{\bQ}
\safemath{\matR}{\bR}
\safemath{\matS}{\bS}
\safemath{\matT}{\bT}
\safemath{\matU}{\bU}
\safemath{\matV}{\bV}
\safemath{\matW}{\bW}
\safemath{\matX}{\bX}
\safemath{\matY}{\bY}
\safemath{\matZ}{\bZ}
\safemath{\matzero}{\bmzero}

\safemath{\matDelta}{\bDelta}
\safemath{\matLambda}{\bLambda}
\safemath{\matPhi}{\bPhi}
\safemath{\matSigma}{\bSigma}
\safemath{\matOmega}{\bOmega}
\safemath{\matTheta}{\bTheta}

\safemath{\matidentity}{\matI}
\safemath{\matone}{\matO}

\safemath{\rnda}{A}
\safemath{\rndb}{B}
\safemath{\rndc}{C}
\safemath{\rndd}{D}
\safemath{\rnde}{E}
\safemath{\rndf}{F}
\safemath{\rndg}{G}
\safemath{\rndh}{H}
\safemath{\rndi}{I}
\safemath{\rndj}{J}
\safemath{\rndk}{K}
\safemath{\rndl}{L}
\safemath{\rndm}{M}
\safemath{\rndn}{N}
\safemath{\rndo}{O}
\safemath{\rndp}{P}
\safemath{\rndq}{Q}
\safemath{\rndr}{R}
\safemath{\rnds}{S}
\safemath{\rndt}{T}
\safemath{\rndu}{U}
\safemath{\rndv}{V}
\safemath{\rndw}{W}
\safemath{\rndx}{X}
\safemath{\rndy}{Y}
\safemath{\rndz}{Z}

\safemath{\rveca}{\bimA}
\safemath{\rvecb}{\bimB}
\safemath{\rvecc}{\bimC}
\safemath{\rvecd}{\bimD}
\safemath{\rvece}{\bimE}
\safemath{\rvecf}{\bimF}
\safemath{\rvecg}{\bimG}
\safemath{\rvech}{\bimH}
\safemath{\rveci}{\bimI}
\safemath{\rvecj}{\bimJ}
\safemath{\rveck}{\bimK}
\safemath{\rvecl}{\bimL}
\safemath{\rvecm}{\bimM}
\safemath{\rvecn}{\bimN}
\safemath{\rveco}{\bomO}
\safemath{\rvecp}{\bimP}
\safemath{\rvecq}{\bimQ}
\safemath{\rvecr}{\bimR}
\safemath{\rvecs}{\bimS}
\safemath{\rvect}{\bimT}
\safemath{\rvecu}{\bimU}
\safemath{\rvecv}{\bimV}
\safemath{\rvecw}{\bimW}
\safemath{\rvecx}{\bimX}
\safemath{\rvecy}{\bimY}
\safemath{\rvecz}{\bimZ}

\safemath{\rvecxi}{\bmxi}
\safemath{\rveclambda}{\bmlambda}
\safemath{\rvecmu}{\bmmu}
\safemath{\rvectheta}{\bmtheta}
\safemath{\rvecphi}{\bmphi}

\safemath{\rmatA}{\bimA}
\safemath{\rmatB}{\bimB}
\safemath{\rmatC}{\bimC}
\safemath{\rmatD}{\bimD}
\safemath{\rmatE}{\bimE}
\safemath{\rmatF}{\bimF}
\safemath{\rmatG}{\bimG}
\safemath{\rmatH}{\bimH}
\safemath{\rmatI}{\bimI}
\safemath{\rmatJ}{\bimJ}
\safemath{\rmatK}{\bimK}
\safemath{\rmatL}{\bimL}
\safemath{\rmatM}{\bimM}
\safemath{\rmatN}{\bimN}
\safemath{\rmatO}{\bimO}
\safemath{\rmatP}{\bimP}
\safemath{\rmatQ}{\bimQ}
\safemath{\rmatR}{\bimR}
\safemath{\rmatS}{\bimS}
\safemath{\rmatT}{\bimT}
\safemath{\rmatU}{\bimU}
\safemath{\rmatV}{\bimV}
\safemath{\rmatW}{\bimW}
\safemath{\rmatX}{\bimX}
\safemath{\rmatY}{\bimY}
\safemath{\rmatZ}{\bimZ}

\safemath{\rmatDelta}{\bimDelta}
\safemath{\rmatLambda}{\bimLambda}
\safemath{\rmatPhi}{\bimPhi}
\safemath{\rmatSigma}{\bimSigma}
\safemath{\rmatOmega}{\bimOmega}
\safemath{\rmatTheta}{\bimTheta}

\usepackage{amssymb}
\usepackage{amsfonts}
\usepackage{mathrsfs}
\usepackage{xspace}
\usepackage{bm}
\usepackage{fancyref}
\usepackage{textcomp}

\usepackage{multirow}
\usepackage{stmaryrd}

\newenvironment{textbmatrix}{	\setlength{\arraycolsep}{2.5pt}%
								\big[\begin{matrix}}{\end{matrix}\big]%
								\raisebox{0.08ex}{\vphantom{M}}}

\def\be{\begin{equation}}
\def\ee{\end{equation}}
\def\een{\nonumber \end{equation}}
\def\mat{\begin{bmatrix}}
\def\emat{\end{bmatrix}}
\def\btm{\begin{textbmatrix}}
\def\etm{\end{textbmatrix}}

\def\ba#1\ea{\begin{align}#1\end{align}}
\def\bas#1\eas{\begin{align*}#1\end{align*}}
\def\bs#1\es{\begin{split}#1\end{split}}
\def\bg#1\eg{\begin{gather}#1\end{gather}}
\def\bml#1\eml{\begin{multline}#1\end{multline}}
\def\bi#1\ei{\begin{itemize}#1\end{itemize}}

\newcommand{\lefto}{\mathopen{}\left}

\DeclareMathOperator*{\argmin}{arg\;min}		
\DeclareMathOperator*{\argmax}{arg\;max}		
\DeclareMathOperator{\Exop}{\opE}			

\newcommand{\Ex}[2]{\ensuremath{\Exop_{#1}\lefto[#2\right]}} 	

\newcommand{\conj}[1]{\ensuremath{#1^{*}}} 	
\newcommand{\tp}[1]{\ensuremath{#1^{T}}} 		
\newcommand{\herm}[1]{\ensuremath{#1^{H}}} 	
\newcommand{\inv}[1]{\ensuremath{#1^{-1}}} 	

\safemath{\dirac}{\delta}					
\safemath{\krond}{\dirac}					

\safemath{\upto}{\uparrow}
\safemath{\downto}{\downarrow}
\safemath{\iu}{j}							
\safemath{\ev}{\lambda}						
\safemath{\hilseqspace}{l^{2}}				
\newcommand{\banachfunspace}[1]{\setL^{#1}}	
\safemath{\hilfunspace}{\banachfunspace{2}}	
\newcommand{\floor}[1]{\lfloor #1 \rfloor}

\safemath{\SNR}{\textit{SNR}} 				
\safemath{\PAR}{\textit{PAR}} 				
\safemath{\No}{N_0}							
\safemath{\Es}{E_s}							
\safemath{\Eb}{E_b}							
\safemath{\EbNo}{\frac{\Eb}{\No}}
\safemath{\EsNo}{\frac{\Es}{\No}}

\DeclareMathOperator{\CHop}{\ensuremath{\opH}} 
\safemath{\tvir}{\rndh_{\CHop}}				
\safemath{\tvtf}{\rndl_{\CHop}}				
\safemath{\spf}{\rnds_{\CHop}}				
\safemath{\bff}{H_{\CHop}}					

\safemath{\ircf}{r_{h}}						
\safemath{\tftvcf}{r_{s}}					
\safemath{\tfcf}{r_{l}}						
\safemath{\bfcf}{r_{H}}						

\safemath{\tcorr}{c_h}						
\safemath{\scf}{c_{s}}						
\safemath{\tfcorr}{c_{l}}					
\safemath{\fcorr}{c_{H}}						

\safemath{\mi}{I}							
\safemath{\capacity}{C}						

\safemath{\normal}{\mathcal{N}}			
\safemath{\jpg}{\mathcal{CN}}			
\safemath{\mchain}{\leftrightarrow}		

\safemath{\dB}{\,\mathrm{dB}}
\safemath{\dBm}{\,\mathrm{dBm}}
\safemath{\Hz}{\,\mathrm{Hz}}
\safemath{\kHz}{\,\mathrm{kHz}}
\safemath{\MHz}{\,\mathrm{MHz}}
\safemath{\GHz}{\,\mathrm{GHz}}
\safemath{\s}{\,\mathrm{s}}
\safemath{\ms}{\,\mathrm{ms}}
\safemath{\mus}{\,\mathrm{\text{\textmu}s}}
\safemath{\ns}{\,\mathrm{ns}}
\safemath{\ps}{\,\mathrm{ps}}
\safemath{\meter}{\,\mathrm{m}}
\safemath{\mm}{\,\mathrm{mm}}
\safemath{\cm}{\,\mathrm{cm}}
\safemath{\m}{\,\mathrm{m}}
\safemath{\W}{\,\mathrm{W}}
\safemath{\mW}{\, \mathrm{mW}}
\safemath{\J}{\,\mathrm{J}}
\safemath{\K}{\,\mathrm{K}}
\safemath{\bit}{\,\mathrm{bit}}
\safemath{\nat}{\,\mathrm{nat}}

\safemath{\define}{\triangleq}			

\providecommand{\inner}[2]{\ensuremath{\langle#1,#2\rangle}}
\safemath{\equivalent}{\sim}
\safemath{\distas}{\sim}					
\safemath{\sdiff}{\Delta}				

\safemath{\reals}{\mathbb{R}}
\safemath{\positivereals}{\reals_{+}}
\safemath{\integers}{\mathbb{Z}}
\safemath{\posint}{\integers_{+}}
\safemath{\naturals}{\mathbb{N}}
\safemath{\posnaturals}{\naturals_{+}}
\safemath{\complexset}{\mathbb{C}}
\safemath{\rationals}{\mathbb{Q}}

\newcommand*{\fancyrefapplabelprefix}{app}		
\newcommand*{\fancyrefthmlabelprefix}{thm}		
\newcommand*{\fancyreflemlabelprefix}{lem}		
\newcommand*{\fancyrefcorlabelprefix}{cor}		
\newcommand*{\fancyrefdeflabelprefix}{def}		
\newcommand*{\fancyrefproplabelprefix}{prop}		
\newcommand*{\fancyrefexmpllabelprefix}{exmpl}
\newcommand*{\fancyrefalglabelprefix}{alg}		
\newcommand*{\fancyreftbllabelprefix}{tbl}		

\frefformat{vario}{\fancyrefseclabelprefix}{Section~#1}
\frefformat{vario}{\fancyrefthmlabelprefix}{Theorem~#1}
\frefformat{vario}{\fancyreftbllabelprefix}{Table~#1}
\frefformat{vario}{\fancyreflemlabelprefix}{Lemma~#1}
\frefformat{vario}{\fancyrefcorlabelprefix}{Corollary~#1}
\frefformat{vario}{\fancyrefdeflabelprefix}{Definition~#1}
\frefformat{vario}{\fancyreffiglabelprefix}{Fig.~#1}
\frefformat{vario}{\fancyrefapplabelprefix}{Appendix~#1}
\frefformat{vario}{\fancyrefeqlabelprefix}{(#1)}
\frefformat{vario}{\fancyrefproplabelprefix}{Proposition~#1}
\frefformat{vario}{\fancyrefexmpllabelprefix}{Example~#1}
\frefformat{vario}{\fancyrefalglabelprefix}{Algorithm~#1}

 \newtheorem{thm}{Theorem}
 \newtheorem{cor}[thm]{Corollary}   
 \newtheorem{prop}{Proposition}

\safemath{\dictab}{[\,\dicta\,\,\dictb\,]}

\safemath{\ysig}{\bmy}
\safemath{\ysighat}{\hat{\ysig}}
\safemath{\ysigdim}{M}
\safemath{\xsig}{\bmx}
\safemath{\xsigdim}{N}
\safemath{\nx}{n_x}
\safemath{\zsig}{\bmz}
\safemath{\zsigdim}{\ysigdim}
\safemath{\rsig}{\bmr}
\safemath{\Adict}{\bA}
\safemath{\Adicttilde}{\widetilde{\Adict}}
\safemath{\Adictdim}{\outputdim\times\xsigdim}
\safemath{\avec}{\bma}
\safemath{\avectilde}{\tilde{\avec}}
\safemath{\Bdict}{\bB}
\safemath{\Bdicttilde}{\widetilde{\Bdict}}
\safemath{\Cdict}{\bC}
\safemath{\cvec}{\bmc}
\safemath{\Ddict}{\bD}
\safemath{\Ddictdim}{\ysigdim\times\xsigdim}
\safemath{\dvec}{\bmd}
\safemath{\Ddicttilde}{\widetilde{\bD}}
\safemath{\Bonb}{\bB}
\safemath{\bvec}{\bmb}
\safemath{\Bonbdim}{\ysigdim\times\ysigdim}
\safemath{\noise}{\bmn}
\safemath{\noisedim}{\ysigim}
\safemath{\err}{\bme}
\safemath{\errdim}{\ysigdim}
\safemath{\errset}{\setE}
\safemath{\nerr}{n_e}
\safemath{\delop}{\bP_\errset}
\safemath{\delopc}{\bP_{{\errset}^c}}

\safemath{\cplxi}{\imath}
\safemath{\cplxj}{\jmath}

\safemath{\dict}{\matD}
\safemath{\inputdim}{N}		
\safemath{\outputdim}{M}		
\safemath{\sparsity}{S}	
\safemath{\inputdimA}{{N_a}}	
\safemath{\inputdimB}{{N_b}}	
\safemath{\elemA}{{n_a}}	
\safemath{\elemB}{{n_b}}	
\safemath{\resA}{\matR_a}	
\safemath{\resB}{\matR_b}	
\safemath{\subD}{\matS} 
\safemath{\subA}{\matS_a} 
\safemath{\subB}{\matS_b} 
\safemath{\dicta}{\matA} 	
\safemath{\dictb}{\matB} 	
\safemath{\hollowS}{H}
\safemath{\hollowA}{H_a}
\safemath{\hollowB}{H_b}
\safemath{\cross}{Z}
\safemath{\coh}{\mu_d}			
\safemath{\coha}{\mu_a}			
\safemath{\cohb}{\mu_b}			
\safemath{\mubs}{\nu}	
\safemath{\cohm}{\mu_m} 
\safemath{\dictset}{\setD}	
\safemath{\dictsetp}{\dictset(\coh,\coha,\cohb)}	
\safemath{\dictsetgen}{\dictset_\text{gen}}
\safemath{\dictsetgenp}{\dictsetgen(\coh)}
\safemath{\dictsetonb}{\dictset_\text{onb}}
\safemath{\dictsetonbp}{\dictsetonb(\coh)}

\safemath{\leftside}{U}
\safemath{\rightsideA}{R_a}
\safemath{\rightsideB}{R_b}

\safemath{\indexS}{\setI_S} 

\safemath{\na}{n_a}			
\safemath{\nb}{n_b}			
\safemath{\coeffa}{p_i}	
\safemath{\coeffb}{q_j}	
\safemath{\seta}{\setP}		
\safemath{\setb}{\setQ}     
\safemath{\setw}{\setW}	
\safemath{\setz}{\setZ}	
\safemath{\cola}{\veca}		
\safemath{\colb}{\vecb}		
\safemath{\cold}{\vecd}		
\safemath{\inputvec}{\vecx} 	
\safemath{\error}{\vece}	
\safemath{\noiseout}{\vecz} 	
\safemath{\inputvecel}{x}
\safemath{\inputveca}{\vecx_a}
\safemath{\inputvecb}{\vecx_b}
\safemath{\outputvec}{\vecy}	
\safemath{\lambdamin}{\lambda_{\mathrm{min}}}

\safemath{\elltwo}{\ell_2}
\safemath{\ellone}{\ell_1}
\safemath{\ellzero}{\ell_0}
\safemath{\ellinf}{\ell_\infty}
\safemath{\ellinftilde}{\ell_{\widetilde\infty}}
\safemath{\licard}{Z(\coh,\coha,\cohb)}
\safemath{\xsol}{\hat{x}}
\safemath{\xbord}{x_b}		
\safemath{\xstat}{x_s}		
\safemath{\xstatLone}{\tilde{x}_s}
\safemath{\order}{\mathcal{O}} 
\safemath{\scales}{\Theta} 
\safemath{\ones}{\mathbf{1}} 
\safemath{\zeroes}{\mathbf{0}} 
\safemath{\thlone}{\kappa(\coh,\cohb)} 
\safemath{\constoneA}{\delta} 
\safemath{\constoneB}{\epsilon} 
\safemath{\nlarge}{L}				   
\safemath{\sumlarge}{S_\nlarge}
\safemath{\maxlarger}{P_\nlarge}	   
\safemath{\Pzero}{\textrm{P0}}	
\safemath{\Pone}{\textrm{P1}}
\safemath{\vecfir}{\vecw}			 
\safemath{\vecsec}{\vecz}
\safemath{\elvecfir}{w}              
\safemath{\elvecsec}{z}				 
\safemath{\nlargefir}{n}
\safemath{\normout}{\gamma}
\safemath{\auxfun}{h}
\safemath{\supp}{\textrm{supp}}

\safemath{\indexa}{\ell}
\safemath{\indexb}{r}
\safemath{\indexc}{i}
\safemath{\indexd}{j}

\safemath{\project}{P}
\usepackage{framed}

\IEEEoverridecommandlockouts
\allowdisplaybreaks 

\newcommand{\startsquarepar}{%
    \par\begingroup \parfillskip 0pt \relax}
\newcommand{\stopsquarepar}{%
    \par\endgroup}

\safemath{\Hj}{\bmj}
\safemath{\sj}{w}
\safemath{\Ej}{E_w}
\safemath{\quant}{Q}
\safemath{\compquant}{\mathcal{Q}}
\safemath{\Cy}{\bC_{\bmy}}

\definecolor{blue(ncs)}{rgb}{0.0, 0.53, 0.74}
\definecolor{bluelink}{HTML}{0000EE}
\definecolor{crabred}{HTML}{BE1831}

\begin{document}

\title{Hybrid Jammer Mitigation for \\ All-Digital mmWave Massive MU-MIMO}

\author{\IEEEauthorblockN{ Gian Marti\textsuperscript{\twemoji{shell},\twemoji{crab}}, Oscar Casta\~neda\textsuperscript{\twemoji{shell},\twemoji{crab}}, 
Sven Jacobsson\textsuperscript{\twemoji{shrimp}},\\
Giuseppe Durisi\textsuperscript{\twemoji{squid}}, Tom Goldstein\textsuperscript{\twemoji{lobster}}, 
Christoph Studer\textsuperscript{\twemoji{crab}}}\\
\textit{\textsuperscript{\twemoji{crab}}ETH Zurich, Zurich, Switzerland; e-mail: gimarti@ethz.ch, caoscar@ethz.ch, studer@ethz.ch}\\
\textit{\textsuperscript{\twemoji{shrimp}}Ericsson Research, Gothenburg, Sweden; e-mail: sven.jacobsson@ericsson.com}\\
\textit{\textsuperscript{\twemoji{squid}}Chalmers University of Technology, Gothenburg, Sweden; e-mail: durisi@chalmers.se}\\
\textit{\textsuperscript{\twemoji{lobster}}University of Maryland, College Park, MD; e-mail: tomg@cs.umd.edu}
\thanks{\textsuperscript{\twemoji{shell}}GM and OC contributed equally to this work.}
\thanks{The work of OC and CS was supported in part by ComSenTer, one of six centers in JUMP, a SRC program sponsored by DARPA. The work of GM, OC, and CS was supported in part by an ETH Research Grant. The work of CS was supported in part by the U.S.\ National Science Foundation (NSF) under grants CNS-1717559 and ECCS-1824379.}
\thanks{\hspace{-2.5mm}Emojis by Twitter, Inc.~and other contributors are licensed under \color{crabred}\href{https://creativecommons.org/licenses/by/4.0/}{CC-BY 4.0}\color{black}.} 
}

\maketitle

\renewcommand{\abstractname}{tl;dr}

\begin{abstract}
Low-resolution analog-to-digital converters (ADCs) simplify the design of millimeter-wave (mmWave) 
massive multi-user multiple-input multiple-output (MU-MIMO) basestations, but increase vulnerability to jamming attacks.
As a remedy, we propose HERMIT (short for Hybrid jammER MITigation), a method that combines a hardware-friendly adaptive analog transform with a corresponding digital equalizer: 
The analog transform removes most of the jammer's energy prior to data conversion; the digital equalizer suppresses jammer residues while detecting the legitimate transmit data.
We provide theoretical results that establish the optimal analog transform as a function of the user equipments' and the jammer's channels.
Using simulations with mmWave channel models, we demonstrate the superiority of HERMIT compared both to purely digital 
jammer mitigation as well as to a recent hybrid method  that mitigates jammer interference with a nonadaptive analog transform.

\end{abstract}

\section{Introduction}

Millimeter-wave (mmWave) massive multi-user multiple-input multiple-output (MU-MIMO) is expected to be a key technology in meeting
the ever-growing demands for increased data rates~\cite{swindlehurst14a, rappaport15a}.
However, a practical deployment of basestations (BSs) with hundreds of antennas presents serious implementation challenges in terms of power consumption, circuit complexity, and hardware cost. 
A promising way of addressing these challenges is to rely on low-resolution analog-to-digital converters (ADCs) at the BS
\cite{heath-jr.15a, roth18a, jacobsson17b}.
Unfortunately, such architectures exhibit increased vulnerability to jamming attacks, which could for instance be enacted by a malicious transmitter or by a malfunctioning user equipment (UE) \cite{marti2021snips}.
 
In principle, attacks during which the jammer's transmit characteristics are known can be mitigated easily and effectively, 
e.g., by digitally projecting the received signal on the subspace orthogonal to the jammer channel \cite{subbaram1993interference, yan14a}. 
Such digital equalization methods assume that the jammer's impact on the digital receive signal amounts to a linear superposition of signals. 
Unfortunately, this is no longer the case if the BS relies on low-resolution ADCs---a strong jammer will either cause the ADCs to clip
or will cause their quantization range to widen, thereby drowning the UEs' signal in quantization noise.
As a result, the effectiveness of linear digital equalization methods that mitigate jamming attacks is significantly reduced~\cite{marti2021snips}.

\vspace{-0.25mm}
\subsection{Contributions}
\vspace{-0.25mm}
In this work, we propose HERMIT (short for Hybrid jammER MITigation), 
a practical method to mitigate strong jamming attacks during
uplink transmission in low-resolution mmWave massive MU-MIMO systems.
HERMIT is a hybrid method that combines (i) an \emph{adaptive} analog transform that removes most of the 
jammer's energy before data conversion 
to shield the low-resolution ADCs from strong jamming signals
with (ii) a digital equalizer that estimates the UE transmit symbols 
while taking into account both the jammer and the analog transform. 
The proposed analog transform is adaptive, as it depends on the UEs' and jammer's channels, and is  designed to be suited to hardware implementation.
We develop theoretical results that establish the optimal parameters for the analog transform as a function
of these channels.
Using simulations with mmWave channel models,
we show that HERMIT is able to mitigate strong jamming attacks  far more effectively than a purely digital method. 
We also show that HERMIT outperforms a recently-proposed hybrid method for jammer mitigation which relies on a 
\textit{nonadaptive}~analog~transform~\cite{marti2021snips}.

\vspace{-0.25mm}
\subsection{Relevant Prior Work}
\vspace{-0.25mm}
Improving the resiliency of MU-MIMO systems against jamming attacks has been the focus of 
considerable attention in recent years. 
Several methods have been developed whose goal is to \textit{detect} jamming attacks~\cite{kapetanovic13a, akhlaghpasand18a, teeti2021one}. 
This allows methods whose goal is to \textit{mitigate} jamming attacks to take their detection for granted (as we will do).
Researchers have studied how to mitigate jamming attacks of various kinds, 
such as attacks with stationary~signature \cite{yan14a, shen14a, do18a, akhlaghpasand20a, akhlaghpasand20b}, as well as attacks
in which the jammer only interferes at specific time instants, e.g., during pilot transmission \cite{kapetanovic13a} 
or during data transmission \cite{yan14a}.
In this work, we focus on the mitigation of stationary jamming attacks. 
To mitigate such attacks on small-scale MIMO systems, the earlier works~\cite{shen14a, yan14a} 
have developed methods that digitally project the receive signals on the subspace
orthogonal to the jammer's channel while using coherent~\cite{yan14a} or noncoherent~\cite{shen14a}
 data transmission.
In the context of massive MIMO, different variants of jammer-resilient linear data detectors have been proposed 
for single-user~\cite{do18a} and multi-user \cite{akhlaghpasand20a, akhlaghpasand20b} systems. 

However, only a few studies have considered the effects of low-resolution ADCs on jammer mitigation. 
In \cite{akhlaghpasand20b}, the~effects of hardware impairments (including quantization
errors) have been modeled as additive Gaussian noise which does not accurately capture the 
effects of low-resolution ADCs.
Reference~\cite{teeti2021one} has devised a method for the detection of jamming attacks
on massive MU-MIMO systems with one-bit ADCs, but not for their mitigation. 
In \cite{pirzadeh2019mitigation}, a method for jammer mitigation that builds upon 
spatial sigma-delta converters has been proposed and analyzed, but for a single UE only.
A hybrid method for jammer mitigation in massive MU-MIMO systems with conventional low-resolution  ADCs has recently been proposed in~\cite{marti2021snips}. 
The method utilizes a \emph{nonadaptive} analog transform, inspired 
by the beamspace transform \cite{brady13}, to focus the jammer's energy onto a subset of ADCs, 
thereby enabling signal detection based on the outputs of the jammer-free ADCs.
In contrast to \cite{marti2021snips}, we propose an \emph{adaptive} method in which the analog transform depends on the UEs' and jammer's channels. 

\subsection{Notation}
Matrices and column vectors are represented by boldface uppercase and lowercase letters, respectively.
For a matrix $\bA$, the conjugate transpose is $\bA^H$, the entry in the $\ell$th row and $k$th column is $[\bA]_{(\ell,k)}$, and the Frobenius norm is $\| \bA \|_F$.
The $N\!\times\!N$ identity matrix is $\bI_N$.
For a vector $\bma$, the $k$th entry~is $a_k$, the $\ell_2$-norm is $\|\bma\|_2$, the real part is $\Re\{\bma\}$, and the imaginary part is $\Im\{\bma\}$.
Furthermore, $\text{diag}(\bma)$ is a diagonal matrix whose diagonal is formed by $\bma$. 
Expectation with respect to the random vector~$\bmx$ is denoted by \Ex{\bmx}{\cdot}.
The floor function $\floor{x}$ returns the greatest integer less than or equal to $x$.
We define $i^2=-1$.

\section{System Model}
We consider the uplink of a mmWave massive MU-MIMO system in which $U$ single-antenna UEs transmit data 
to a \mbox{$B$-antenna} BS while a single-antenna jammer emits a stationary jamming signal
that interferes with the receive signal at the~BS.
The involved channels are assumed to be frequency-flat, and we model transmission as 
\begin{align}
	\bmy = \bH\bms + \Hj\sj + \bmn. \label{eq:ant_io}
\end{align}
Here, $\bmy\in\complexset^B$ is the receive vector at the BS, $\bH\in\complexset^{B\times U}$ models the MIMO uplink channel matrix, $\bms\in\setS^U$ is the vector whose entries correspond to the per-UE transmit symbols and take value in some constellation set $\setS$ (e.g., $16$-QAM), $\Hj\in\complexset^B$ is the channel from the jammer to the BS, $\sj\in\complexset$ is the jammer transmit signal, and $\bmn\in\complexset^B$ is i.i.d.\ circularly-symmetric complex Gaussian noise with per-entry variance~$\No$.

\begin{figure}[tp]
\centering
\includegraphics[width=\columnwidth]{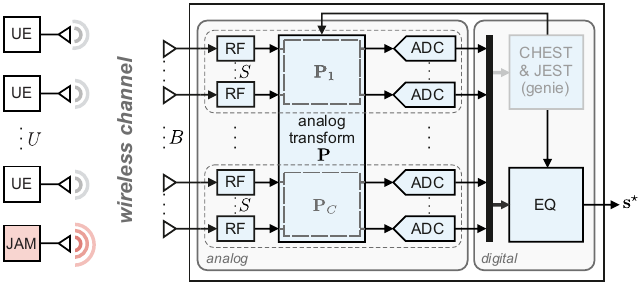}
\caption{System overview of HERMIT: The $B$ radio-frequency (RF) chains can be grouped into $C$ clusters of
size $S$. The analog transform $\bP$ can then operate clusterwise on the analog baseband signals to remove 
most of the jammer's energy prior to A/D conversion. A digital equalization step then detects the data while
cancelling the jammer signal's residues.}
\label{fig:system_overview}
\end{figure}

HERMIT, our hybrid jammer mitigation method, consists of three stages (\fref{fig:system_overview}):
An analog transform $\bP$ operating on the unquantized receive signal\footnote{This 
means that $\bP$ also transforms the noise term $\bmn$ in \eqref{eq:ant_io}. 
If $\bmn$ is given its usual interpretation of representing thermal noise, then, in 
our model, thermal noise arises prior to the analog transform only. In reality, 
thermal noise would arise before, during, and after the analog transform stage. 
We use this simplified model in order to not overcomplicate things and because the 
relevant performance-limiting factor in this work is at any rate not the thermal noise, but the jammer interference.
} 
according to $\bmy_\bP = \bP\bmy,$
followed by analog-to-digital (A/D) conversion, modelled as a quantization step
$\bmr = \compquant(\bmy_\bP),$ 
and, finally, a digital equalization step that estimates the symbol vector $\bms^\star = \bW \bmr.$
Section~\ref{sec:analog_proj} is devoted to the analog transform; Section~\ref{sec:hermit}
provides the details of HERMIT.

\startsquarepar 
In the remainder, we assume that the UE transmit symbols $s_u$, $u=1,\dots,U$, are independent  zero-mean with variance $\Es$, so that $\Ex{\bms}{\bms\herm{\bms}}=\Es\bI_U$.
The jamming signal $\sj$ is modelled as circularly-symmetric complex Gaussian with variance $\Ej$,
and probabilistic quantities are assumed to be mutually independent. 
We assume throughout the paper that the channels $\bH$ and $\Hj$, as well as 
the quantities $\Es,\Ej$, and $\No$, are known at the BS. 
\stopsquarepar 

\section{Analog Transforms for Jammer Inhibition} \label{sec:analog_proj}
In order to stop the jammer from unduly widening the ADCs' quantization range, the analog transform $\bP$ should remove as much of the jammer's power as possible. 
For this reason, we want $\bP$ to minimize the mean squared error (MSE) between the transformed signal $\bP\bmy$
and the jammer-free signal $\bH\bms+\bmn$, 
\begin{align}
	\widehat\bP = \argmin_{\bP\in\setP} \Ex{\bms,\sj,\bmn}{\|\bP\bmy - (\bH\bms + \bmn)\|^2}.  \label{eq:opt1}
\end{align}
If $\setP$ were simply the set of all complex-valued $B\times B$-matrices, then 
the solution of this optimization problem would be
\begin{align}
\widehat\bP = \left(\Es\bH\herm{\bH} \!+\! \No \bI_B \right)\!\inv{\left(\Es\bH\herm{\bH} \!+\!\Ej\Hj\herm{\Hj} \!+\! \No\bI_B \right)}\!.
\end{align}
However, such an operator $\widehat\bP$ is ill-suited for analog implementation, 
since it requires $B^2$ complex-valued multiplications for computing $\widehat\bP\bmy$. 
To obtain an operator $\bP$ that is hardware-friendly, we thus restrict the set $\setP$ to matrices which 
(i)~do not result in a dense matrix-vector-product when calculating $\bP\bmy$, 
(ii)~have entries that can be represented with low-resolution numbers, 
and (iii)~enable decentralized computation of $\bP\bmy$.

\subsection{Structured Jammer Inhibition}
To avoid a dense matrix-vector-product with $B^2$ complex multiplications,
we first constrain $\setP$ to all matrices of the form
\begin{align}
	\bP = \bI_B - \beta\bmb\herm{\bma}, \label{eq:pmat_struc}
\end{align}
where we demand that $\bma\in\setA^B, \bmb\in\setB^B$, with $\setA\subset\opC$ and $\setB\subset\opC$, and with $\beta\in\mathbb{C}$.
This operator structure is inspired by~\cite{castaneda19fame} and with it, $\bP\bmy$ can be calculated as 
\begin{align}
	\bP\bmy &= \bmy - \beta (\herm{\bma}\bmy) \bmb, \label{eq:proj_explicit}
\end{align}
i.e., we only need to calculate the inner product $\herm{\bma}\bmy$
and subtract an accordingly scaled version of the vector $\bmb$ from~$\bmy$.

To bring out the dependence on $\beta, \bmb, \bma$, we reformulate~\eqref{eq:opt1}~as
\begin{align}
	\{\hat\beta, \hat\bmb, \hat\bma \}= \argmin_{\beta\in\mathbb{C},\bmb\in\setB^B,  \bma\in\setA^B}
	\Ex{\bms,\sj,\bmn}{\|\beta \bmb\herm{\bma}\bmy - \bmj\sj\|^2}.  \label{eq:opt2}
\end{align}

\begin{prop}\label{prop:proj}
The optimization problem \eqref{eq:opt2} is separable in~$\bmb$ and $\bma$, and its solution is given by
\begin{align}
	\hat\bmb &= \argmax_{\bmb\in\setB^B} \frac{|\herm{\Hj}\bmb|^2}{\|\bmb\|^2}, \label{eq:optb} \\
	\hat\bma &= \argmax_{\bma\in\setA^B} \frac{|\herm{\Hj}\bma|^2}{\herm{\bma}\Cy\bma}, \label{eq:opta} \\
	\hat\beta &= \frac{\Ej \herm{\Hj}\hat\bma\herm{\hat\bmb}\Hj}{\|\hat\bmb\|^2 \herm{\hat\bma}\Cy\hat\bma}, \label{eq:optbeta}
\end{align}	
where 
\begin{align}
\Cy = \Es\bH\herm{\bH} + \Ej\Hj\herm{\Hj} + \No \bI_B
\end{align}
is the covariance matrix of $\bmy$.
\end{prop}
\begin{cor}\label{cor:unconstrained}
If we drop the alphabet constraints\footnote{We use the subscript $\mathbb{C}$ to 
denote the absence of any alphabet constraints.} and instead have $\setA=\setB=\mathbb{C}$, 
then a solution to \eqref{eq:opt2} is\footnote{Rescaled versions of \eqref{eq:sol_unconst} that lead to \eqref{eq:p_unconst} are also solutions of \eqref{eq:opt2}.}
\begin{align}
\hat\bmb_{\mathbb{C}} = \Hj,\quad
\hat\bma_{\mathbb{C}} = \Ej\inv{\Cy}\Hj,\quad
\hat\beta_{\mathbb{C}}=1, \label{eq:sol_unconst}
\end{align}
and we have 
\begin{align}
	\widehat\bP_{\mathbb{C}} = \bI_B - \Ej \Hj\herm{\Hj}\inv{\Cy}. \label{eq:p_unconst}
\end{align}
\end{cor}
The proofs for Proposition \ref{prop:proj} and Corollary \ref{cor:unconstrained} are relegated to the appendices. 
The solution in \eqref{eq:sol_unconst} has an elegant interpretation when plugged into \eqref{eq:proj_explicit}:
The inner product $\herm{\hat\bma_{\opC}}\bmy$ computes the linear minimum MSE (LMMSE) estimate $\sj_{\text{LMMSE}}$ of $\sj$ based on $\bmy$.
In effect, \eqref{eq:proj_explicit} can be restated~as 
\begin{align}
	\widehat\bP_{\mathbb{C}} \bmy = \bmy - \Hj \sj_{\text{LMMSE}}.
\end{align}

\subsection{Transforms with Finite Alphabets}
\begin{figure}[tp]
\centering
\includegraphics[width=0.87\columnwidth]{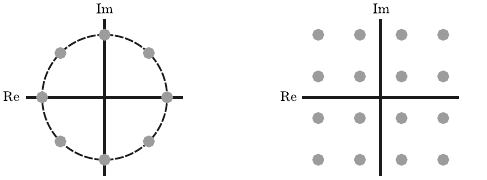}
\caption{Illustration of a phase quantization alphabet of cardinality $8$ (left), 
and of a quadrature quantization alphabet of cardinality $16$ (right).}
\label{fig:pesklis_quemlis}
\end{figure}

As outlined so far, the transform $\bP$ avoids a dense matrix-vector-product, 
but its entries potentially need high-resolution numbers to be adequately represented.
To eliminate this requirement, we now restrict the sets $\setB$ and $\setA$ to low-cardinality finite alphabets.
We consider two types of alphabets that are particularly suited to analog implementation (\fref{fig:pesklis_quemlis}): 
Alphabets with values of constant modulus that are equidistantly spaced in phase (as in phase-shift keying constellations) 
and alphabets with values that are equidistantly spaced on a square grid (as in quadrature amplitude modulation constellations).
We refer to the use of these alphabets as \textit{phase quantization} and \textit{quadrature quantization}, respectively. 
Most works on analog beamforming networks favor phase quantization, as it can be implemented with simple phase shifters~\cite{venkateswaran10analog,sohrabi16}.
Quadrature quantization, however, has higher expressivity, enabling control of the amplitude as well as of the phase, 
and is also practically viable~\cite{castellanos2018hybridamp,naviasky21isscc}.
Both alphabets are fully specified by their \textit{alphabet cardinality (AC)}: their scale is arbitrary 
as it can be absorbed into~$\beta$; see~\eqref{eq:pmat_struc} and \eqref{eq:optbeta}.

With such finite alphabets, solving \eqref{eq:optb} and \eqref{eq:opta} is difficult in general.
We thus resort to only finding approximations \mbox{$\tilde\bmb\in\setB^B$} and \mbox{$\tilde\bma\in\setA^B$} to the optimal solution $\hat\bmb$ and $\hat\bma$.
In this paper, we approximate these vectors simply by computing the 
optimal unconstrained values $\hat\bmb_{\mathbb{C}}$ and $\hat\bma_{\mathbb{C}}$ as in \eqref{eq:sol_unconst} and 
then quantizing them componentwise. Mathematically, this can be expressed~as
\begin{align}
	\tilde\bmb = \argmin_{\bmb\in\setB^B} \|\bmb - \hat\bmb_{\mathbb{C}}\|^2,\quad
	\tilde\bma = \argmin_{\bma\in\setA^B} \|\bma - \hat\bma_{\mathbb{C}}\|^2, \label{eq:ba_FA}
\end{align}
with $\hat\bmb_{\mathbb{C}},\,\hat\bma_{\mathbb{C}}$ given by \eqref{eq:sol_unconst}, 
leading to 
\begin{align}
\widetilde\bP = \bI_B - \hat\beta\tilde\bmb\herm{\tilde\bma}, \label{eq:P_FA}
\end{align}
with $\hat\beta$ as in \eqref{eq:optbeta}.\footnote{Even if $\tilde\bmb$ and $\tilde\bma$ 
differ from the values in \eqref{eq:optb} and \eqref{eq:opta}, the optimal value of $\beta$ is still given 
by \eqref{eq:optbeta} (with $\tilde\bmb$ and $\tilde\bma$ in lieu of $\hat\bmb$ and $\hat\bma$), see \eqref{eq:general_opt_beta}.}
How to approximate \eqref{eq:optb} and \eqref{eq:opta} directly using more sophisticated algorithms will be discussed in 
\cite{marti2021fajita}.

\subsection{Decentralized Transforms}
By now, only one obstacle to the practicality of the analog transform remains:  
The computation of $\bP\bmy$ involves inputs from $B$ antennas, and such global 
operations are difficult to implement in analog.
We therefore decentralize the transform into~$C$ clusters of size
$S=B/C$: The set $\setP$ then corresponds to all matrices of block-diagonal structure 
\begin{align}
\bP = \text{diag}(\bP_1, \dots, \bP_C), \label{eq:clusters}
\end{align}
where the blocks $\bP_c$ are of the form 
	$\bP_c = \bI_S - \beta_c\bmb_c\herm{\bma_c}.$
The block-diagonal structure of $\bP$ implies that $\bP\bmy$ can be calculated clusterwise. 
Also, denoting $\bmy = \tp{[\tp{\bmy_1}, \dots, \tp{\bmy_C}]}$ (and similarly for the jammer
channel $\Hj$ and the $S\!\times\!S$ diagonal blocks of the covariance matrix $\Cy$), the optimization 
problem in~\eqref{eq:opt2} can be decomposed into the individual clusters: The solution for the 
$c$th cluster is obtained from Proposition \ref{prop:proj} by simply replacing the dimension $B$ with $S$ 
and inserting the subscript~$c$.

\section{HERMIT: Hybrid Jammer Mitigation}
\label{sec:hermit}

\subsection{The Analog Transform Stage}
HERMIT is a hybrid method for jammer mitigation that consists of an
analog transform as in Section~\ref{sec:analog_proj}, data conversion, and digital equalization (\fref{fig:system_overview}). 
In the analog transform stage, the unquantized receive vector $\bmy$ is transformed~as 
\begin{align}
	\bmy_\bP = \widetilde\bP \bmy, \label{eq:analog_transform}
\end{align}
where $\widetilde\bP$ is decentralized as in 
\eqref{eq:clusters},  
and where its blocks $\widetilde\bP_c$ are finite-alphabet transforms obtained from 
\eqref{eq:ba_FA} and \eqref{eq:P_FA}.

\subsection{Data Conversion}
To accurately model the quantization errors imposed by the low-resolution data-converters, 
we assume that the transformed signal $\bmy_\bP$ is quantized as
\begin{equation}
	\bmr = \inv{\bG}\left(\quant\left(\Re\{\bG \bmy_\bP\}\right) + i\quant\left(\Im\{\bG \bmy_\bP\}\right)\right)\!. \label{eq:quantization1}
\end{equation}
Here, $\bG=\text{diag}(g_1,\dots,g_B)$ is a diagonal matrix which we introduce to represent gain control of the ADCs, 
and whose values we will choose with the aim of normalizing the quantizer's input range (see below).
The quantization function~$Q(\cdot)$ operates entrywise on its input and represents a $q$-bit uniform midrise 
quantizer with step size $\Delta$ defined as 
\begin{equation} \label{eq:quantizer}
\quant(x) \define
    \begin{cases}
      \Delta\floor{\frac{x}{\Delta}}+\frac{\Delta}{2}, & \text{if}\ |x|\leq \Delta 2^{q-1}\\
      \frac{\Delta}{2}(2^q-1)\frac{x}{|x|}, & \text{if}\ |x|>\Delta 2^{q-1}.
    \end{cases}
\end{equation}
We set the step size $\Delta$ to the value that minimizes the MSE between the quantizer's input $x$ 
and its output $Q(x)$ when assuming that $x$ is standard normal \cite{max60a}. For simplicity, we henceforth denote 
\eqref{eq:quantization1} simply by 
\begin{equation}
	\bmr = \compquant(\bmy_\bP). \label{eq:quantization2}
\end{equation}

This quantization procedure introduces errors that are~correlated with the quantizer's input. 
We use Bussgang's decomposition \cite{bussgang52a} to take these errors into account in the equalization step (see below).
Bussgang's decomposition allows~us to decompose the quantization of a real-valued statistical signal~$x$~as
\begin{align}
	Q(x) = \gamma\,x + d.
\end{align}
Here, the constant $\gamma$ is the quantizer's \emph{Bussgang gain}, 
and the distortion $d$ is a zero-mean random variable that is uncorrelated with $x$. 
The Bussgang gain and variance of the distortion are 
\begin{equation}
	\gamma = \frac{\Ex{}{\quant(x)x}}{\Ex{}{x^2}}	 
\end{equation}
and
\begin{equation}
	D = \Ex{}{d^2} = \Ex{}{\quant(x)^2} - \gamma^2 \Ex{}{x^2}, \label{eq:bussgang_distortion}
\end{equation}
respectively.
In order to apply Bussgang's decomposition to our quantization procedure, we assume
that the components of the quantizer's input are Gaussian with zero mean and unit variance. 
Given \eqref{eq:ant_io}, the assumption of zero mean is true, 
and the assumption of unit variance can be met by choosing the entries of the gain control matrix $\bG$
as follows:
\begin{align}
	g_k = \sqrt{\frac{2}{[\bP\Cy\herm{\bP}]_{(k,k)}}}.
\end{align} 
We can then rewrite \eqref{eq:quantization2} as 
\begin{align}
	\bmr &= \compquant(\bmy_\bP) \\
	&= \inv{\bG}\left(\quant\left(\Re\{\bG \bmy_\bP\}\right) + i\quant\left(\Im\{\bG \bmy_\bP\}\right)\right)\\
	&= \inv{\bG}\left(\gamma\,\Re\{\bG \bmy_\bP\} + \bmd_r + i\left(\gamma\,\Im\{\bG \bmy_\bP\} + \bmd_i\right)\right) \\
	&= \gamma\, \inv{\bG} \left(\Re\{\bG \bmy_\bP\} + i\Im\{\bG \bmy_\bP\}\right) +\inv{\bG}( \bmd_r + i\bmd_i) \\
	&= \gamma\, \bmy_\bP + \inv{\bG} \bmd, \label{eq:bussgang_final}
\end{align}	
where we define $\bmd = \bmd_r + i\bmd_i$.
We make the idealizing assumption that the covariance matrix of $\bmd$ can
be approximated~as
\begin{equation}
	\bC_\bmd = \Ex{\bmd}{\bmd\herm{\bmd}} \approx 2D\,\bI_B,\label{eq:assump_diag}
\end{equation}
where the values on the diagonal follow from \eqref{eq:bussgang_distortion}.

\subsection{Digital Equalization}
Combining equations \eqref{eq:ant_io}, \eqref{eq:analog_transform}, and \eqref{eq:bussgang_final}, we can rewrite the 
\mbox{analog-input} to digital-output relation of our system model~as 
\begin{align}
	\bmr = \gamma\widetilde\bP \left(\bH\bms + \Hj\sj + \bmn \right) + \inv{\bG}\bmd.
\end{align}
Given our assumptions about the distributions of the jammer input $\sj$, the thermal noise $\bmn$, 
and the quantization error $\bmd$, the LMMSE estimate of $\bms$
is 
\begin{align}
	\bms^\star = \bW \bmr,
\end{align}
where the matrix $\bW$ is given by
\begin{align}
\bW = \frac{1}{\gamma}\herm{\bH}\herm{\widetilde\bP}\Big(\widetilde\bP\bH\herm{\bH}\herm{\widetilde\bP}
&+ \frac{\Ej}{\Es}\widetilde\bP\Hj\herm{\Hj}\herm{\widetilde\bP} \nonumber\\ 
+ \frac{\No}{\Es}\widetilde\bP\herm{\widetilde\bP} &+ \frac{2D}{\gamma^2\Es}\bG^{-2}\Big)^{\!\!-1}\!\!.
\end{align}

\section{Results}

\begin{figure*}[tp]
\centering
\subfigure[LoS]{
\includegraphics[width=0.315\linewidth]{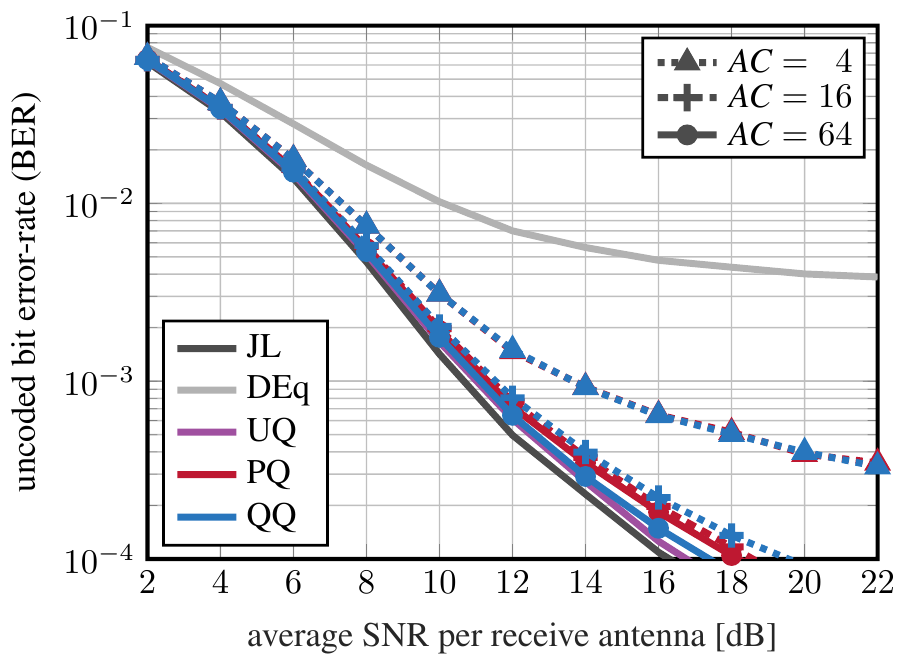}
\label{fig:ac_sweep:los}
}
\subfigure[non-LoS]{
\includegraphics[width=0.315\linewidth]{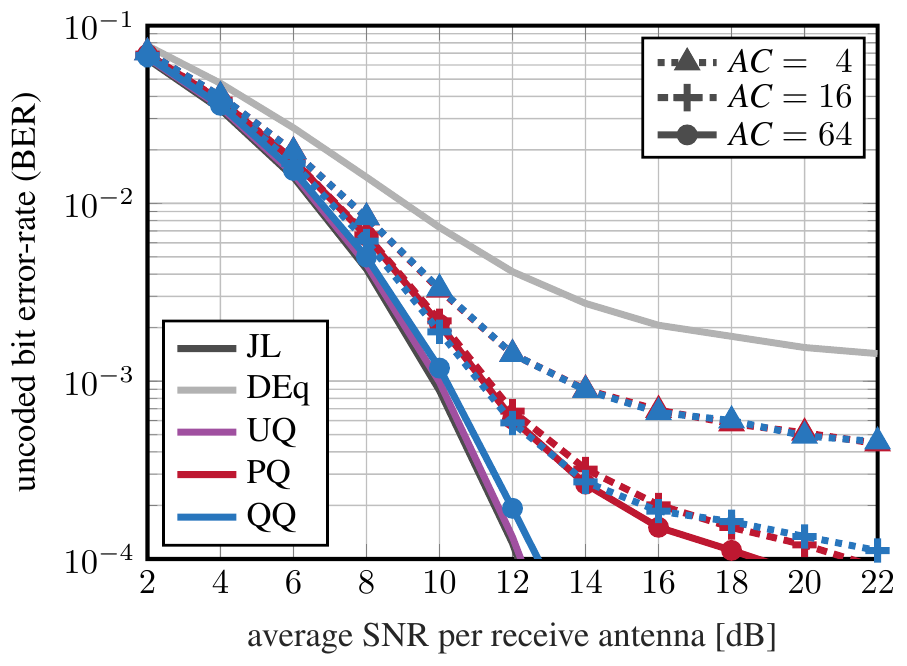}
\label{fig:ac_sweep:nlos}
}
\subfigure[LoS]{
\includegraphics[width=0.315\linewidth]{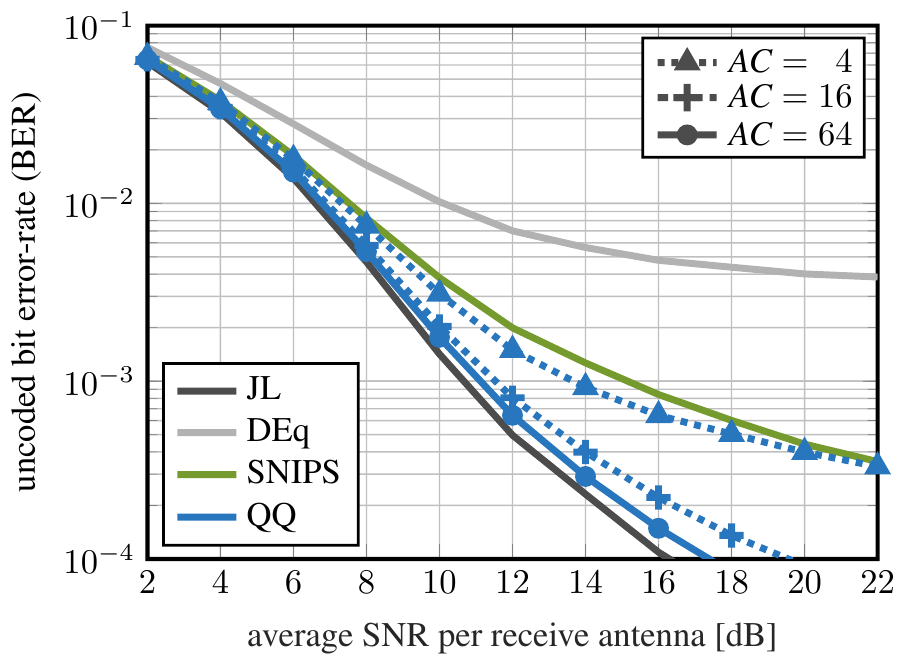}
\label{fig:ac_sweep:snips}
}
\caption{Uncoded bit error-rate (BER) of different jammer mitigation methods when varying the alphabet cardinality 
(\textit{AC}) of HERMIT. The relative jammer power is $\rho=25$\,dB, the ADC resolution is $q=4$\,bits, 
and the cluster size of HERMIT and SNIPS is $S=64$.}
\label{fig:ac_sweep}
\end{figure*}
\begin{figure*}[tp]
\centering
\subfigure[LoS]{
\includegraphics[width=0.315\linewidth]{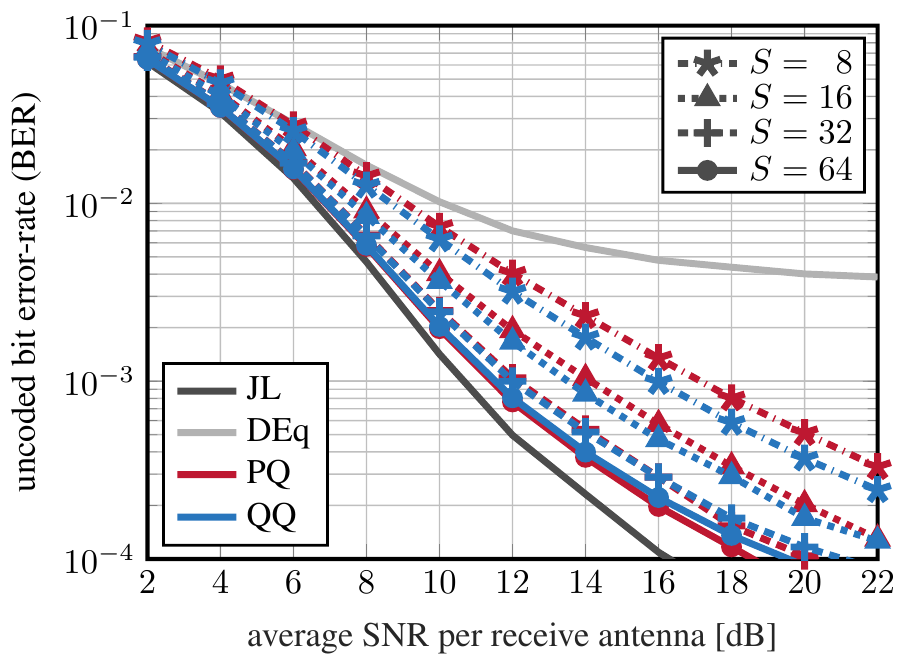}
\label{fig:s_sweep:los}
}
\subfigure[non-LoS]{
\includegraphics[width=0.315\linewidth]{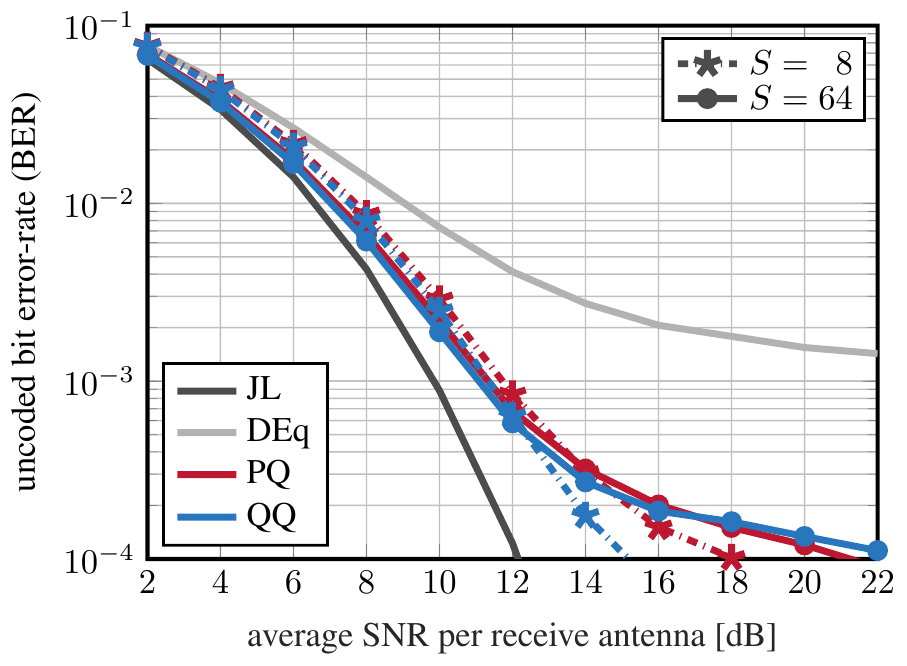}
\label{fig:s_sweep:nlos}
} 
\subfigure[LoS]{
\includegraphics[width=0.315\linewidth]{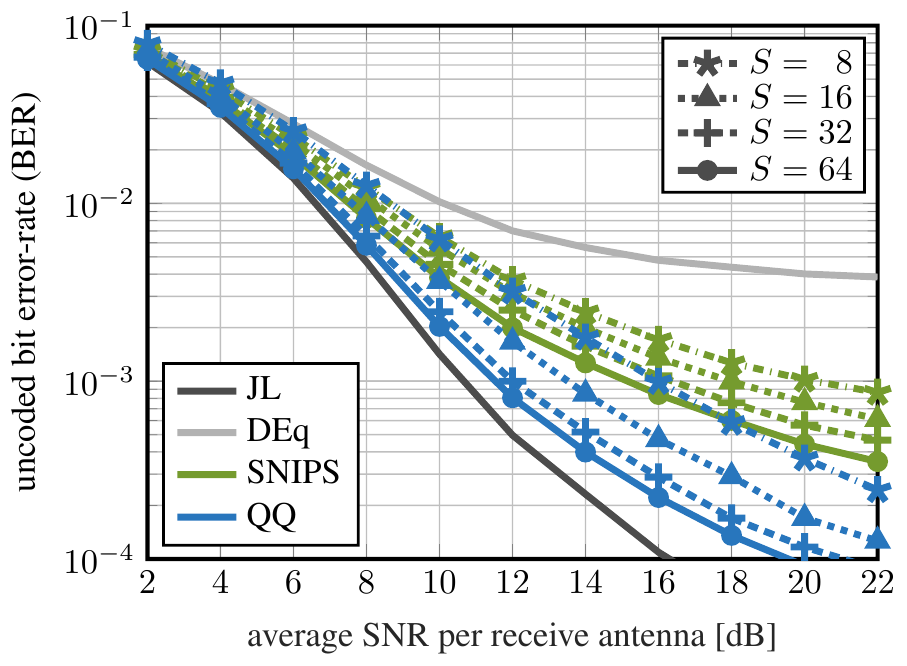}
\label{fig:s_sweep:snips}
}
\caption{Uncoded bit error-rate (BER) of different jammer mitigation methods when varying the cluster size $S$ 
of HERMIT and SNIPS. The relative jammer power is $\rho=25$\,dB, the ADC resolution is $q=4$\,bits, 
and the alphabet cardinality of HERMIT is $\textit{AC}=16$.}
\label{fig:s_sweep}
\vspace{-0.17cm}
\end{figure*}

\subsection{Simulation Setup}
We simulate a mmWave massive MU-MIMO system in which $U=32$ single-antenna UEs transmit data to a BS 
with \mbox{$B=256$} antennas in the presence of a single-antenna jammer.
We create LoS and non-LoS channels using the QuaDRiGa mmMAGIC urban microcellular (UMi) 
model~\cite{jaeckel2014quadriga}. We use a carrier frequency of 60\,GHz and a uniform linear array (ULA) with 
antennas spaced at half a wavelength. The $U$~UEs and the jammer are randomly placed at distances 
between~10\,m and 100\,m within a 120$^\circ$ sector in front of the BS. The minimum angular
separation between any two UEs, as well as between any UE and the jammer, is 1°. 
The UEs transmit 16-QAM symbols and the jamming signal $\sj$ is circularly-symmetric complex
Gaussian noise with variance $\Ej$. We assume $\pm3$\-dB per-UE power control, so that 
the ratio between maximum and minimum per-UE receive power is $4$. 
To display our results, we define the average receive signal-to-noise ratio (SNR) as
\begin{align}
\textit{SNR} \define \frac{\Ex{\bms}{\|\bH\bms\|_2^2}}{\Ex{\bmn}{\|\bmn\|_2^2}}.
\end{align}
To quantify how the strength of the jammer interference compares to the strength of the UE signals, 
we define the relative (to a single UE) jammer power $\rho$ as  
\begin{align}
\rho \define \frac{U\Ex{\sj}{\|\Hj\sj\|_2^2}}{\Ex{\bms}{\|\bH\bms\|_2^2}} =  \frac{U\Ej \|\Hj\|_2^2}{\Es\|\bH\|_F^2}.
\end{align}
Our performance metric is the uncoded bit error-rate (BER).
We evaluate the performance of HERMIT when quantized in phase~(PQ) or in quadrature~(QQ); 
as a reference, we also include the performance of HERMIT without quantization~(UQ), i.e., with $\widetilde\bP$  given by 
\eqref{eq:p_unconst}.
We also simulate a jammerless (JL) system that serves as a lower bound on the BER, as well as two other jammer mitigation schemes that serve as baselines: (i) a purely digital equalizer (DEq) that follows the data pipeline from \fref{sec:hermit} without incorporating any kind of analog transform (i.e., $\widetilde\bP=\bI$) and (ii) the nonadaptive method SNIPS~\cite{marti2021snips} whose beam-slicing analog transform does not require knowledge of the jammer's channel.

\subsection{Simulation Results}
\begin{figure*}[tp]
\centering
\subfigure[$q=4$\,bits]{
\includegraphics[width=0.315\linewidth]{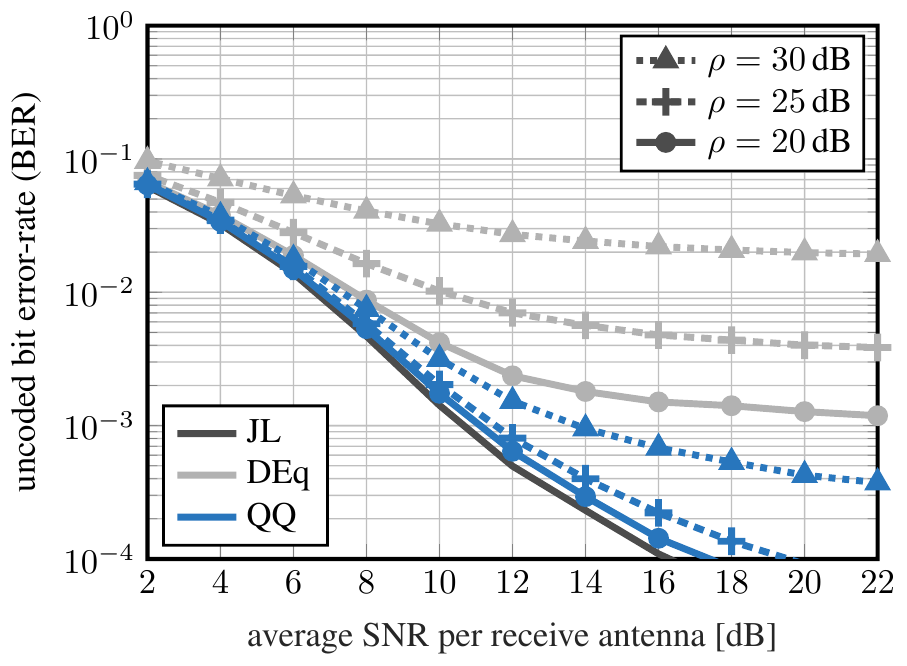}
\label{fig:rho_sweep:4bit}
}
\subfigure[$q=3$\,bits]{
\includegraphics[width=0.315\linewidth]{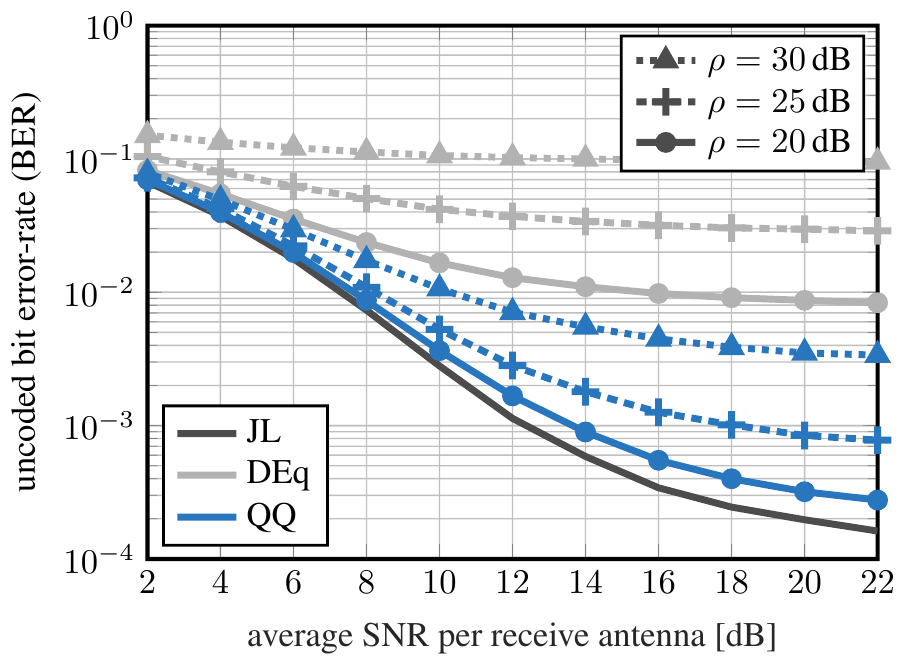}
\label{fig:rho_sweep:3bit}
}
\subfigure[$q=4$\,bits]{
\includegraphics[width=0.315\linewidth]{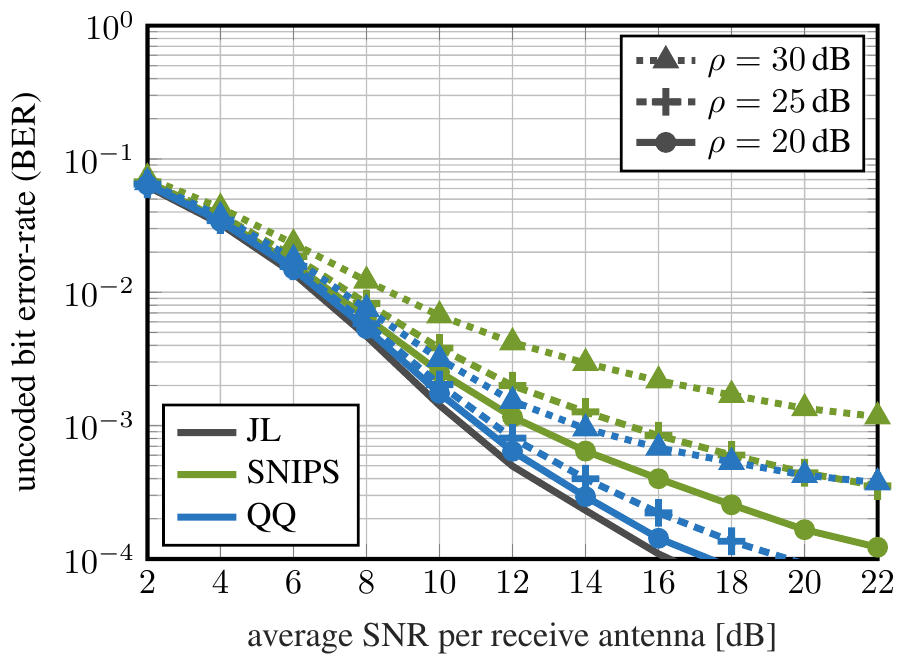}
\label{fig:rho_sweep:snips}
}
\caption{Uncoded bit error-rate (BER) of different jammer mitigation methods when varying the ADC resolution $q$
and the relative jammer strength $\rho$. The alphabet cardinality of HERMIT is $\textit{AC}=16$, 
and the cluster size of HERMIT and SNIPS is $S=64$.}
\label{fig:rho_sweep}
\end{figure*}

We evaluate the performance of HERMIT while varying the propagation conditions (LoS/non-LoS), the alphabet cardinality $AC$, the cluster size $S$, the relative jammer power $\rho$, and the ADC resolution $q$.
\fref{fig:ac_sweep} shows the BER for the different jammer-mitigation methods when using 4-bit ADCs while facing a $\rho=25$\,dB jammer under LoS (Figs.~\ref{fig:ac_sweep:los}, \ref{fig:ac_sweep:snips}) or non-LoS (\fref{fig:ac_sweep:nlos}) conditions, with HERMIT (and SNIPS, \fref{fig:ac_sweep:snips}) using cluster size $S=64$.
Evidently, the purely digital equalizer DEq (unaided by an analog transform) is unable to mitigate the jammer effectively, with an error floor as high as $0.4\%$ BER (\fref{fig:ac_sweep:los}). 
In stark contrast, \mbox{HERMIT-UQ} approaches the interference-free JL performance within 0.5\,dB.

\textit{1) What alphabet cardinality is needed?}
We start by analyzing the impact of the alphabet cardinality under LoS propagation, as shown in \fref{fig:ac_sweep:los}.
With \textit{AC} as small as 4, HERMIT significantly outperforms DEq (regardless of whether PQ or QQ is used), but there is still an SNR loss of 2.7\,dB at 0.1\% BER compared to UQ.
This gap closes rapidly for increased \textit{AC}, with $\textit{AC}=64$ 
approaching the UQ performance at 0.1\% BER within 0.1\,dB~for QQ and 0.3\,dB for PQ.

\fref{fig:ac_sweep:nlos} shows the influence of \textit{AC} under non-LoS propagation.
As in LoS, HERMIT outperforms DEq already for $\textit{AC}=4$ (both PQ and QQ) and can in fact, for UQ, meet the JL performance.
Unlike the LoS case, however, $\textit{AC}=16$ is not enough to approach the UQ performance.
Moreover, increasing \textit{AC} from 16 to 64 yields a significant performance improvement only for QQ, which approaches the UQ performance within 0.5\,dB, whereas a significant gap persists for PQ.

\fref{fig:ac_sweep:snips} again shows results for LoS channels, but includes the performance of SNIPS.
For simplicity, only the QQ version of HERMIT is shown.
While SNIPS clearly improves over the DEq baseline, it is outperformed by HERMIT already when~$\textit{AC}=4$, 
illustrating the advantage of using~an analog transform that adapts to the UEs' and jammer's channels.

\textit{2) The impact of cluster size:}
\fref{fig:s_sweep} shows the effect of the cluster size $S$ on the performance of HERMIT.
The ADC resolution is 4 bits and the jammer has a relative power \mbox{$\rho=25$\,dB}, while HERMIT uses PQ and QQ alphabets with 
$\textit{AC}=16$. 
\fref{fig:s_sweep:los} shows the BER of HERMIT for LoS propagation as the cluster size is varied from $S=8$ to $S=64$.
HERMIT outperforms the DEq baseline already for the smallest considered cluster size and its performance increases further with $S$.
We note that, although the performance of PQ and QQ is largely similar, QQ has a slight advantage for $S\!<\!32$, while PQ exhibits a slight advantage at high SNR for $S\!\geq\!32$.

For \mbox{non-LoS} propagation, the impact of $S$ is quite different, as shown by \fref{fig:s_sweep:nlos}: 
Increasing $S$ from 8 to 64 does not significantly improve the performance---on the contrary, the high-SNR performance gets worse as $S$ increases.
We attribute this phenomenon to the reduced expressivity of the scaling factors $\beta_1,\dots,\beta_C$: Systems with larger cluster size have fewer clusters $C$, and hence fewer factors $\beta_c$ for adjusting the scale of the analog transform.
This reinforces our previous observation that amplitude information is crucial for non-LoS propagation.

\fref{fig:s_sweep:snips} again compares HERMIT to SNIPS, whose cluster size can also be varied.
The performance of both HERMIT and SNIPS increases with $S$, and at low-to-medium SNR, 
SNIPS with large clusters can in fact outperform HERMIT with $S=8$. 
However, at high SNR, HERMIT always outperforms SNIPS regardless of cluster size. 
Furthermore, HERMIT consistently outperforms SNIPS with equally sized clusters.

\textit{3) Considering different levels of relative distortion:}
\fref{fig:rho_sweep} shows results for different levels of relative distortion at the ADCs, i.e., 
for different combinations of jammer strengths~$\rho$ and ADC resolutions $q$.
For simplicity, we focus on LoS propagation, and on HERMIT with QQ, $\textit{AC}=16$, and \mbox{$S=64$}.
\fref{fig:rho_sweep:4bit} illustrates the impact of the relative jammer power~$\rho$ when using 4-bit ADCs,
where HERMIT always outperforms the DEq baseline by a significant margin: Even in the strongest $\rho=30$\,dB jammer scenario, 
HERMIT achieves a better performance than DEq does in the weakest $\rho=20$\,dB jammer scenario.
Moreover, for the $\rho=20$\,dB jammer, the performance of HERMIT with $\textit{AC}=16$ is virtually identical to the JL performance, exhibiting only a 0.5\,dB SNR loss at 0.1\% BER.

\fref{fig:rho_sweep:3bit} illustrates the impact of the relative jammer power~$\rho$ when using 3-bit ADCs, which in trend is 
similar to the 4-bit case, although the gap between HERMIT and JL widens.
Here, a particularly interesting scenario is that of the $\rho=30$\,dB jammer: While DEq has an error floor of 10\% BER at 10\,dB SNR, HERMIT achieves tenfold lower BER at the same SNR.

\fref{fig:rho_sweep:snips} compares HERMIT against SNIPS for 4-bit ADCs.
For the same relative jammer power $\rho$, HERMIT always outperforms SNIPS and its performance for a \mbox{$\rho=25$\,dB} jammer is better than that of SNIPS for a $\rho=20$\,dB jammer.

In summary, these results show that HERMIT can effectively mitigate strong jammers over a wide range of parameters, 
with a performance superior to the one of purely digital~jammer mitigation 
or of hybrid methods with nonadaptive analog~transforms.
The practicality of HERMIT is supported through these experiments: They show that HERMIT can be decentralized into 
clusters of $32$ or $64$ antennas while maintaining excellent performance 
and that PQ with 16 elements is enough to mitigate jammers under LoS conditions, 
although~non-LoS conditions may require QQ with higher-order alphabets.

\section{Conclusion}

We have proposed HERMIT, a novel hybrid method for the mitigation of strong jammers in low-resolution mmWave massive \mbox{MU-MIMO} BSs.
HERMIT consists of an adaptive analog transform that removes most of the jammer's energy prior to A/D conversion 
and of a mean-square-optimal digital equalizer that detects the UE transmit symbols while cancelling the residues of the
jammer's interference.
The optimal value of this analog transform, which relies on finite-alphabets and is decentralized to facilitate practical implementation, 
is determined by theoretical results that we have established.

We have demonstrated the efficacy of HERMIT through extensive simulations on mmWave channel models. 
Our results show that HERMIT (i) can mitigate jammers with small losses in performance when compared to  jammerless scenarios 
and (ii) outperforms hybrid methods whose analog transform does not adapt to  
the users' and jammer's channels.

\appendices
\section{Proof of Proposition \ref{prop:proj}}
We can rewrite the optimization objective as 
\begin{align}
&\hspace{-0.5cm}\Ex{}{\|\beta \bmb\herm{\bma}\bmy - \bmj\sj\|^2}\\
&= |\beta|^2 \|\bmb\|^2 \herm{\bma}\Ex{}{\bmy\herm{\bmy}}\bma + \|\Hj\|^2\Ex{}{|\sj|^2}\nonumber\\
&\hphantom{=}~ - \beta\herm{\Hj}\bmb\herm{\bma}\Ex{}{\bmy \conj{\sj}} - \conj{\beta}\Ex{}{\sj\herm{\bmy}}\bma\herm{\bmb}\Hj\\
&=  |\beta|^2 \|\bmb\|^2 \herm{\bma}\Cy\bma + \Ej \|\Hj\|^2 \nonumber\\
&\hphantom{=}~	- \beta\Ej\herm{\Hj}\bmb\herm{\bma}\Hj  - \conj{\beta}\Ej\herm{\Hj}\bma\herm{\bmb}\Hj. \label{eq:opt3}
\end{align}
This objective is quadratic in $\beta$. Taking the Wirtinger derivative with respect to $\beta$ and equating with zero gives
\begin{align}
	\beta = \frac{\Ej \herm{\Hj}\bma\herm{\bmb}\Hj}{\|\bmb\|^2 \herm{\bma}\Cy\bma}. \label{eq:general_opt_beta}
\end{align}	
Plugging this value back into \eqref{eq:opt3} and dropping the \mbox{$\Ej \|\Hj\|^2$-term} that does not depend on
$\bma$ and $\bmb$ yields
\begin{align}
	-\Ej^2 \frac{|\herm{\Hj}\bma\herm{\bmb}\Hj|^2}{\|\bmb\|^2 \herm{\bma}\Cy\bma} 
	= - \Ej^2\, \frac{|\herm{\Hj}\bmb|^2}{\|\bmb\|^2} \, \frac{|\herm{\Hj}\bma|^2}{\herm{\bma}\Cy\bma}. \label{eq:opt4}
\end{align}
It is evident that \eqref{eq:opt4} can be minimized by independently maximizing the factors 
\begin{align}
	\frac{|\herm{\Hj}\bmb|^2}{\|\bmb\|^2} \quad\text{and}\quad \frac{|\herm{\Hj}\bma|^2}{\herm{\bma}\Cy\bma}.
\end{align}

\section{Proof of Corollary \ref{cor:unconstrained}}
We prove that \eqref{eq:sol_unconst} is the solution to \eqref{eq:optb}--\eqref{eq:optbeta} 
when $\setA=\setB=\mathbb{C}$:
That \mbox{$\bmb=\Hj$} maximizes \eqref{eq:optb} directly follows from the Cauchy-Schwarz inequality, which states that 
\begin{align}
	\frac{|\herm{\Hj}\bmb|^2}{\|\bmb\|^2} \leq \frac{\|\Hj\|^2 \|\bmb\|^2 }{\|\bmb\|^2} = \|\Hj\|^2, 
\end{align}
where equality holds if and only if $\bmb$ is colinear with $\Hj$. 
As to the optimal value of $\bma$, we define $\bar{\Hj}\triangleq\inv{\Cy}\Hj$, so we can rewrite
\begin{align}
	\frac{|\herm{\Hj}\bma|^2}{\herm{\bma}\Cy\bma} 
	= \frac{|\herm{\bar\Hj}\Cy\bma|^2}{\herm{\bma}\Cy\bma}
	= \frac{|\inner{\bma}{\bar\Hj}_{\Cy}|^2}{\|\bma\|_{\Cy}^2},
\end{align}
where we denote by $\inner{\cdot}{\cdot}_{\Cy}$ the inner product with respect to the positive definite matrix $\Cy$, 
and by $\|\cdot\|_{\Cy}$ the norm induced by that inner product. 
We can now again use the Cauchy-Schwarz inequality to bound
\begin{align}
	\frac{|\inner{\bma}{\bar\Hj}_{\Cy}|^2}{\|\bma\|_{\Cy}^2} \leq \|\bar\Hj\|_{\Cy}^2
\end{align}
with equality if and only if $\bma$ is colinear with $\bar\Hj = \inv{\Cy}\Hj$.

\balance

\end{document}